
\input harvmac.tex


\Title{KYUSHU-HET-9}
{The massive Schwinger model with $SU(2)_{f}$ on the light cone}


\centerline{Koji Harada\footnote{$^1$}{f77453a@kyu-cc.cc.kyushu-u.ac.jp},
Takanori Sugihara, Masa-aki Taniguchi and Masanobu Yahiro$^\dagger$}
\bigskip
\centerline{Department of Physics, Kyushu University}
\centerline{Fukuoka, 812 JAPAN}
\smallskip
\centerline{$^\dagger$Shimonoseki University of Fisheries}
\centerline{Shimonoseki, 759-65 JAPAN}


\vskip 0.5cm
The massive Schwinger model with two flavors is studied in the strong
coupling region by using light-front Tamm-Dancoff approximation. The mass
spectrum of the lightest particles is obtained numerically. We find that
the mass of the lightest isotriplet (``pion'') behaves as $m^{0.50}$ for
the strong couplings, where $m$ is the fermion mass. We also find that the
lightest isosinglet is not in the valence state (``eta'') which is much
heavier in the strong coupling region, but can be interpreted as a bound
state of two pions. It is 1.762 times heavier than pion at
$m=1.0\times10^{-3}(e/\sqrt\pi)$, while Coleman predicted that the ratio is
$\sqrt3$ in the strong coupling limit. The ``pion decay constant'' is
calculated to be 0.3945.

\Date{9/93}



\newsec{Introduction}

The light-front Tamm-Dancoff (LFTD) approximation has attracted much
attention
recently as an alternative non-perturbative method to lattice
theory\ref\lftd{R. J. Perry, A. Harindranath, and K. G. Wilson, Phys. Rev.
Lett.
{\bf 65} (1990) 2959}. It is the Tamm-Dancoff
approximation\ref\td{I. Tamm, J. Phys. (USSR) {\bf 9} (1945) 449\semi S. M.
Dancoff,
Phys. Rev {\bf 78} (1950) 382\semi H. A. Bethe and F. de Hoffman, {\sl Mesons
and
Fields} (Row, Peterson, Evanston, 1955) Vol. II \semi E. M. Henley and W.
Thirring,
{\sl Elementary Quantum Field Theory} (McGraw-Hill, New York, 1962)} applied
to field
theory quantized on the light cone\ref\light{An extensive list of references
on
light-front physics by A. Harindranath (light.tex) is available via anonymous
ftp
from public.mps.ohio-state.edu under the subdirectory
tmp/infolight.}\ref\easy{S. J.
Brodsky, G. McCartor, H. C. Pauli and S. S. Pinsky, Particle World {\bf 3}
(1993)
109}. The light-cone quantization provides a cure for one of the most serious
problems of the Tamm-Dancoff approximation; in the application of the
Tamm-Dancoff
approximation, one must first specify the ground state, while in the
light-cone
quantization the ground state is relatively simple because the Fock vacuum is
an
eigenstate of the light-front Hamiltonian. The LFTD field theory is a very
attractive and efficient numerical method for relativistic bound
state problems and is intuitively appealing because it is based on
diagonalization of Hamiltonians with the eigenstates being wave functions
for bound states.

There are however several problems in LFTD field theory; (1)
(non-perturbative) renormalization, (2) spontaneous symmery breaking (or
``zero-mode
problem''), and (3) recovery of rotational symmetry. These problems are very
important
in the development of LFTD field theory. We think, however, that it is useful
to see
how far the LFTD field theory goes by studying simple models for
which we can circumvent these problems.

In this paper, we study the massive Schwinger model with 2 flavors in
the strong coupling region non-perturbatively in the LFTD approximation.
Because it is a two-dimensional model, there is no renormalization
problem. \nref\nong{S. Coleman, Commun. Math. Phys. {\bf 31} (1973) 259}
Because the fermions are massive, spontaneous symmetry breaking
does not come into trouble\ref\fone{In two dimensions, spontaneous breaking
of a global continuous symmetry does not occur\nong. We exclude the
possibility of Higgs phenomena too. Note that
 Higgs phenomenon {\it does} occur in the massless Schwinger model, while it
does
not in the massive Schwinger model.}; the global
$SU(2)_A$ symmtry is explicitly (and softly) broken. Because there are no
transverse directions, the rotational symmetry is not broken
from the outset. As we will see, in the strong coupling region, the
structure of the mass spectrum is relatively simple.

The massive Schwinger model\ref\cjs{S. Coleman, R. Jackiw
and L. Susskind, Ann. Phys. {\bf 93} (1975) 267}\ref\coleman{S. Coleman,
Ann. Phys. {\bf 101} (1976) 239} is a generalization of the massless
Schwinger model\ref\schw{J. Schwinger,Phys. Rev. {\bf 128} (1962)
2425}\ref\ls{J. Lowenstein and A. Swieca, Ann. Phys.{\bf68} (1971)
172}\ref\aar{E. Abdalla, M. C. B. Abdalla, and K. Rothe, {\sl
Non-Perturbative Methods in 2 Dimensional Quantum Field Theory} (World
Scientific, Singapore, 1991) and references therein} with the massive
fermion. Both of them have been playing a unique role as simple toy models
for $QCD$\ref\ks{ J. Kogut and L. Susskind, Phys. Rev. {\bf D11} (1975)
3594}. Although the massless Schwinger model is exaclty solvable, the
massive model is no longer exactly solvable. One has to employ some
non-perturbative methods, such as bosonization and Monte-Carlo
simulations\ref\grady{For the massive Schwinger model with $SU(2)_f$, See
M. Grady, Phys. Rev. {\bf D35} (1987) 1961}, to {\it solve} it. In his
beautiful paper\coleman, Coleman studied the massive Schwinger model and
its extention with $SU_f(2)$ flavor (isospin) symmetry by using
bosonization technique.  Among important results, he found the
following things for the model with $SU(2)_f$ symmetry in the strong
coupling limit.  (i) The model is equivalent to the sine-Gordon theory
with $\beta=\sqrt{2\pi}$.  (ii) The lightest particle is an isotriplet,
and the next lightest is an isosinglet.   (iii) The isosinglet/isotriplet
mass ratio is $\sqrt3$.  (iv) The isotriplet is $I^{PG}=1^{-+}$, while the
isosinglet is $0^{++}$, not $0^{--}$. He confessed that he did not
understand why it is so. In this paper, we examine these results
numerically, in the light-front Tamm-Dancoff approximation including up to
4-body states. We do not only confirm his results, but also obtain several
new results. Our main results are:  (i) We also find that the lightest
particle is an isotriplet and the next is an isosinglet in the strong
coupling limit. The isotriplet may be called as a ``pion'', because the
valence component is dominant.  (ii) We calculate the ``pion'' mass as a
function of the fermion mass numerically. It behaves like $m^{0.5007(2)}$ in
the strong coupling limit.
(iii) We calculate the isosinglet/isotriplet mass ratio for various
values of the fermion mass, and find that it is 1.762 for
$m=1.0\times10^{-3}(e/\sqrt\pi)$. (iv) We argue that the lightest
isosinglet is a bound state of two ``pions''. This is our answer to
Coleman's question.  The valence state (``eta'') is much heavier due to the
annihilation force.   (v) We find no eta-eta, pi-eta, pi-pi
in the isotriplet, nor isoquintet bound states in
the strong coupling region.  (vi) We calculate the ``pion dacay constant''
to be 0.3945. (Of course this is merely a two-dimensional analog and has no
physical importance; the ``pions'' do not decay. What we would really like
to do is to demonstrate that we can calculate such phenomenological
quantities like this from the fundamental field theory.)

The massive
Schwinger model in the LFTD approximation has been studied by
Bergknoff\ref\berg{H. Bergknoff, Nucl.Phys. {\bf B122} (1977) 215} and Mo and
Perry\ref\moperry{Y. Mo and R. J. Perry, The Ohio State Univ. preprint May
(1992)}. Our work is based on these works, especially on that of Mo and
Perry.
We refer the readers to them. There are also some papers on the
massive Schwinger model in the discretized light-cone quantization
(DLCQ)\ref\epb{T. Eller, H. C. Pauli and S. Brodsky, Phys, Rev. {\bf D35}
(1987) 1493\semi T. Eller and H. C. Pauli, Z. Phys. {\bf C42} (1989)
59}\ref\YH{ C. M. Yung and C. J. Hamer, Phys. Rev. {\bf D} (1991)
2598}\ref\ltly{ F.
Lenz, M. Thies, S. Levit, and K. Yazaki, Ann. Phys. {\bf 208} (1991)
1}\ref\fujita{ A.
Ogura, T.Tomachi, and T. Fujita, Nihon University preprint NUP-A-92-7\semi
See also T.
Tomachi and T. Fujita, Ann. Phys. {\bf 223} (1993) 197}, which is closely
related to
the LFTD approximation.

The paper is organized as follows: In sec. 2, we will describe the
light-cone quantized massive Schwinger model with $SU(2)_f$ and its
Tamm-Dancoff approximation. The states are classified by the exact flavor
(isospin) symmetry. We truncate the states, keeping only 2- and 4-body
components. The inclusion of 4-body states is essential for our work. As we
will see, the lowest isosinglet state is not in its valence state and
therefore cannot be found in the lowest approximation (up to 4-body states).
In sec. 3, we will describe our numerical method. The method is a
generalization of that of Mo and Perry. We expand the wave functions in
terms of basis functions, which satisfy assumed symmetries under exchanges
of momenta and boundary conditions, and diagonalize the matrices for the
``norm'' and the light-cone Hamiltonian. The ``norm'' is necessary because
the basis functions are not orthonormal. We then give results. The hope for
the LFTD approximation is that the components with large  number of
constituents are suppressed at least for the low-lying states. We will see
that it is the case (except for the lightest isosinglet), by comparing the
calculations with 2- and 4-body states with those with only 2-body states. We
will also show that a small number of basis functions is sufficient to
produce quite accurate values. We will then identify states, by examining
the wave functions. It is a peculiar feature of the LFTD approximation that
we can utilize detailed information of the wave functions. It is also
exploited in calculating the ``pion decay constant''. Sec. 4 is devoted to
discussions. Appendices are collections of conventions, lengthy expressions
and some formulae.

\newsec{Light-Front Tamm-Dancoff Approximation}

\subsec{The massive Schwinger model with two flavors}

In this section we will establish the conventions and provide the readers
with some basic formulae.

The massive Schwinger model is two-dimensional $QED$ with massive fermions.
The
model discussed in this paper involves a flavor $SU(2)_f$ (isospin) symmetry,

\eqn\lag{
{\cal L}=-{1\over 4}F_{\mu\nu}F^{\mu\nu}+
\sum_{i=1}^2\bar\psi_i[\gamma^\mu(i\partial_\mu-eA_\mu)-m]\psi_i,
}
where $i (i=1,2)$ is the label for flavors. Because it
is regarded as a toy model for $QCD$, we sometimes use the words such as
``quark'', ``pion'', and ``eta''. They should not be confused with {\it
real} ones, of course. The equations of motion in the $A^+=0$ gauge
\eqn\eom{
\eqalign{
-\partial_-^2A^-&=\sqrt2e\sum^2_{i=1}\psi^\dagger_{iR}\psi_{iR},\cr
\partial_+\partial_-A^-&=\sqrt2e\sum^2_{i=1}\psi^\dagger_{iL}\psi_{iL},\cr
i\sqrt2\partial_-\psi_{iL}&=m\psi_{iR},\cr
i\sqrt2\partial_+\psi_{iR}&=m\psi_{iL}+\sqrt2eA^-\psi_{iR},\cr
}
}
show that $A^-$ and $\psi_{iL}\equiv\ha(1-\gamma^5)\psi_i$ are dependent
variables, and that only $\psi_{iR}\equiv\ha(1+\gamma^5)\psi_i$ is
independent. (See Appendix A for notation.) These dependent variables
can be eliminated,
\eqn\constr{
\eqalign{
\psi_{iL}(x^-)&=-i{m\over2\sqrt{2}}\int
dy^-\epsilon(x^--y^-)\psi_{iR}(y^-),\cr
A^-(x^-)&=-{e\over\sqrt{2}}\int dy^-\vert x^--y^-\vert
\sum^2_{i=1}\psi^\dagger_{iR}\psi_{iR}(y^-).\cr
}
}
(There can be $x^-$ independent background electric field, which is related
to
the vacuum angle, as Coleman discussed\coleman. We will not consider this
parameter at all in this paper and concentrate only on the
case  $\theta=0$.)

Canonical quantization is performed by assuming the following
anti-commutation
relations for only independent variables.
\eqn\etc{
\eqalign{
\{\psi_{iR}(x),\psi^\dagger_{jR}(y)\}_{x^+=y^+}
&={\delta_{ij}\over\sqrt{2}}\delta(x^--y^-)\cr
\{\psi_{iR}(x),\psi_{jR}(y)\}_{x^+=y^+}
&=\{\psi^\dagger_{iR}(x),\psi^\dagger_{jR}(y)\}_{x^+=y^+}=0.\cr
}
}
In terms of independent variables the Lagrangian can be written as
\eqn\lagg{
\eqalign{
L=&\int dx^-{\cal L}
=i\sqrt2\int dx^- \sum^2_{i=1}:\psi^\dagger_{iR}\partial_+\psi_{iR}:\cr
&+{im^2\over2\sqrt2}\int dx^-dy^-
\sum^2_{i=1}\psi^\dagger_{iR}(x^-)\epsilon(x^--y^-)\psi_{iR}(y^-)\cr
&+{e^2\over4}\int dx^-dy^-
\sum^2_{i=1}j^+(x^-)\vert x^--y^-\vert j^+(y^-).\cr
}
}

In order to construct a well-defined quantum theory, we have to restrict
ourselves
to the $Q=\int dx^- j^+(x^-)=0$ subspase, where the conserved $U(1)_V$
current is
defined by \eqn\vc{
j^\mu=\sum^2_{i=1}:\bar\psi_i\gamma^\mu\psi_i:,
}
otherwise we would have infinite energy.

There are also (anomalous) $U(1)_A$ current
\eqn\avc{
\eqalign{
j_5^\mu&=\sum^2_{i=1}:\bar\psi_i\gamma^\mu\gamma_5\psi_i:,\cr
\partial_\mu j_5^\mu&=2im\sum^2_{i=1}:\bar\psi_i\gamma_5\psi_i:
     +{2e\over\pi}\epsilon^{\mu\nu}\partial_\mu A_\nu,\cr
}
}
as well as $SU(2)_V$ and $SU(2)_A$ currents
\eqn\suc{
\eqalign{
j^{a\mu}&=\sum^2_{i,j}:\bar\psi_i\gamma^\mu(T^a)_{ij}\psi_j:,\cr
j_5^{a\mu}&=\sum^2_{i,j}:\bar\psi_i\gamma^\mu\gamma_5(T^a)_{ij}\psi_j:,\cr
\partial_\mu j^{a\mu}&=0,\cr
\partial_\mu
j_5^{a\mu}&=2im\sum^2_{i,j}:\bar\psi_i\gamma_5(T^a)_{ij}\psi_j:,\cr
}
}
where $T^a=\sigma^a/2$. In the above the normal-ordering is defined with
respect
to the free field expansion;
\eqn\expa{
\psi_{iR}(x^-)={1\over2^{1/4}}\int^\infty_0{dk^+\over2\pi\sqrt{k^+}}
[b_i(k^+)e^{-ik^+x^-}+d^\dagger_i(k^+)e^{ik^+x^-}]
}
where, from \etc,
\eqn\etac{
\{b_i(k^+),b^\dagger_j(l^+)\}=\{d_i(k^+),d^\dagger_j(l^+)\}=2\pi
k^+\delta_{ij}\delta(k^+-l^+).
}
By substituting \expa, the light-cone Hamiltonian $P^{-}$
is written, in terms of these creation and annihilation operators, as
\eqn\ham{
\eqalign{
P^-=&P^-_{free}+P^-_{self}+P^-_0+P^-_2,\cr
P^-_{free}=&{m^2\over4\pi}\sum^2_{i=1}\int^\infty_0{dk\over k^2}
            [b_i^\dagger(k)b_i(k)+d_i^\dagger(k)d_i(k)],\cr
P^-_{self}=&{e^2\over8\pi^2}\sum^2_{i=1}\int^\infty_0{dk_1\over k_1}
            [b_i^\dagger(k_1)b_i(k_1)+d_i^\dagger(k_1)d_i(k_1)]\cr
            &\times\int^\infty_0 dk_2\Big[{1\over(k_1-k_2)^2}
                         -{1\over(k_1+k_2)^2}\Big],\cr
P^-_0=&{e^2\over8\pi^3}\sum^2_{i,j}
       \int^\infty_0{\prod_i^4dk_i\over\sqrt{k_1k_2k_3k_4}}
       \Big\{\big[b_i^\dagger(k_1)b_j^\dagger(k_2)b_j(k_3)b_i(k_4)\cr
       &+d_i^\dagger(k_1)d_j^\dagger(k_2)d_j(k_3)d_i(k_4)\big]
                         {\delta(k_1+k_2-k_3-k_4)\over 2(k_1-k_4)^2}\cr
      &-b_i^\dagger(k_1)d_j^\dagger(k_2)d_j(k_3)b_i(k_4)
                         {\delta(k_1+k_2-k_3-k_4)\over (k_1-k_4)^2}\cr
       &+b_i^\dagger(k_1)d_i^\dagger(k_2)d_j(k_3)b_j(k_4)
                         {\delta(k_1+k_2-k_3-k_4)\over(k_1+k_2)^2}\Big\},\cr
P^-_2=&{e^2\over8\pi^3}\sum^2_{i,j}
        \int^\infty_0{\prod^4_i dk_i\over\sqrt{k_1k_2k_3k_4}}
       \Big[b_i^\dagger(k_1)b_j^\dagger(k_2)d_j^\dagger(k_3)b_i(k_4)\cr
       &+b_i^\dagger(k_4)d_j(k_3)b_j(k_2)b_i(k_1)
        +d_i^\dagger(k_1)d_j^\dagger(k_2)b_j^\dagger(k_3)d_i(k_4)\cr
       &+d_i^\dagger(k_4)b_j(k_3)d_j(k_2)d_i(k_1)\Big]
                         {\delta(k_1+k_2+k_3-k_4)\over(k_1-k_4)^2}.\cr
}
}
Because of light-cone kinematics, the momentum integrations are restricted to
$[0,\infty)$. It explains why the Hamiltonian does not contain the terms
which consists of only creation operators or only annihilation operators;
they would break momentum conservation. Only the exception is the ``zero
modes'',
$k^+=0$. They are in general supposed to be responsible for non-trivial
structure of vacua, such as spontaneous symmetry breaking. In the present
case, however, we have tentatively
dropped them because the presence of the mass term forces the wave
functions vanish at $k^+=0$.

\subsec{ Isospin multiplets}
States are classified by the irreducible representations of the isospin (
$SU(2)_V$ ) symmetry with the conserved isospin charge,
\eqn\iso{
I^a=\int dx^-j^{a+}(x^-).
}

Let us first count how many independent wave functions there are. It is
trivial to see that there are a triplet and a singlet for 2-body states,
$2^2=3+1$. For 4-body states, we have $2^4=16$ states. An elementary
consideration tells that there are one quintet and 3 triplets and 2
singlets, $2^4=5+3\times3+2\times1$. Each multiplet has one independent
wave function. Explicilty, \eqn\twostate{
\eqalign{
\vert 2,2>=&{1\over2}\int^{\cal P}_0{\prod^4_i
dk_i\over(2\pi)^2\sqrt{k_1k_2k_3k_4}}
               \delta(\sum^4_{i=1}k_i-{\cal P})\cr
              &\times\psi_4(k_1,k_2,k_3,k_4)

b_1^\dagger(k_1)b_1^\dagger(k_2)d^\dagger_2(k_3)d^\dagger_2(k_4)\vert0>,\cr
}
}
\eqn\onestate{
\eqalign{
\vert 1,1>=&\int^{\cal P}_0{dk_1dk_2\over2\pi\sqrt{k_1k_2}}
              \delta(k_1+k_2-{\cal
P})\psi_2(k_1,k_2)b_1^\dagger(k_1)d_2^\dagger(k_2)\vert0>\cr
             &+{1\over2}\int^{\cal P}_0{\prod^4_i
dk_i\over(2\pi)^2\sqrt{k_1k_2k_3k_4}}
               \delta(\sum^4_{i=1}k_i-{\cal P})\cr
             &\times\Big\{
              \psi^A(k_1,k_2,k_3,k_4)

[b_1^\dagger(k_1)b_2^\dagger(k_2)d^\dagger_2(k_3)d^\dagger_2(k_4)\cr
             &\qquad \qquad +b_1^\dagger(k_1)b_1^\dagger(k_2)
                 d^\dagger_1(k_3)d^\dagger_2(k_4)]\cr
             &+\sqrt2\psi^{1S}(k_1,k_2,k_3,k_4)

b_1^\dagger(k_1)b_2^\dagger(k_2)d^\dagger_2(k_3)d^\dagger_2(k_4)\cr
             &+\sqrt2\psi^{2S}(k_1,k_2,k_3,k_4)

b_1^\dagger(k_1)b_1^\dagger(k_2)d^\dagger_1(k_3)d^\dagger_2(k_4)
                \Big\}\vert0>,\cr
}
}
\eqn\zerostate{
\eqalign{
\vert 0,0>=&\int^{\cal P}_0{dk_1dk_2\over2\pi\sqrt{k_1k_2}}
            \delta(k_1+k_2-{\cal P})\cr
           &\times\psi_2'(k_1,k_2){1\over\sqrt2}

[b_1^\dagger(k_1)d_1^\dagger(k_2)+b_2^\dagger(k_1)d_2^\dagger(k_2)]\vert0>\cr
             &+{1\over2\sqrt3}\int^{\cal P}_0{\prod^4_i
dk_i\over(2\pi)^2\sqrt{k_1k_2k_3k_4}}
               \delta(\sum^4_{i=1}k_i-{\cal P})\cr
             &\times\Big\{
              \psi_3(k_1,k_2,k_3,k_4)

[b_1^\dagger(k_1)b_1^\dagger(k_2)d^\dagger_1(k_3)d^\dagger_1(k_4)\cr
             &+

2b_1^\dagger(k_1)b_2^\dagger(k_2)d^\dagger_1(k_3)d^\dagger_2(k_4)

+b_2^\dagger(k_1)b_2^\dagger(k_2)d^\dagger_2(k_3)d^\dagger_2(k_4)]\cr
             &+2\sqrt3\psi_0(k_1,k_2,k_3,k_4)

b_1^\dagger(k_1)b_2^\dagger(k_2)d^\dagger_1(k_3)d^\dagger_2(k_4)
                \Big\}\vert0>,\cr
}
}
where ${\cal P}$ is the total light-cone momentum. Other states are obtained
by
the application of $I^-$. Note that the $I=2$ states have no 2-body
components. The
wave functions are assumed to have the following symmetries under exchanges
of
momenta.
\eqn\sym{
\eqalign{
\psi_4(1,2,3,4)=&-\psi_4(2,1,3,4)=-\psi_4(1,2,4,3),\cr
\psi^A(1,2,3,4)=&-\psi^A(2,1,3,4)=-\psi^A(1,2,4,3),\cr
\psi^{1S}(1,2,3,4)=&\psi^{1S}(2,1,3,4)=-\psi^{1S}(1,2,4,3),\cr
\psi^{2S}(1,2,3,4)=&-\psi^{2S}(2,1,3,4)=\psi^{2S}(1,2,4,3),\cr
\psi_3(1,2,3,4)=&-\psi_3(2,1,3,4)=-\psi_3(1,2,4,3),\cr
\psi_0(1,2,3,4)=&\psi_0(2,1,3,4)=\psi_0(1,2,4,3),\cr
}
}
Charge conjugation invariance leads to further restrictions;
\eqn\charge{
\eqalign{\psi_2(1,2)&=\pm\psi_2(2,1),\quad \psi_2'(1,2)=\mp\psi_2'(2,1),\cr
         \psi_4(1,2,3,4)&=\pm\psi_4(3,4,1,2),\quad
\psi^A(1,2,3,4)=\mp\psi^A(3,4,1,2),\cr
       \psi^{1S}(1,2,3,4)&=\pm\psi^{2S}(3,4,1,2),\quad
\psi_0(1,2,3,4)=\pm\psi_0(3,4,1,2),\cr
                \psi_3(1,2,3,4)&=\pm\psi_3(3,4,1,2),\cr
}
}
where the upper (lower) sign coresponds to charge conjugation even (odd). In
the
following, we will not exploit these restrictions but rather use them as
an important check for the results.

\subsec{Einstein-Schr\"odinger equations}
The LFTD approximation is to diagonalize the light-front
Einstein-Schr\"odinger equation,
\eqn\es{
2P^-P^+\vert \psi>=M^2\vert\psi>,
}
in the truncated Fock space. $M^2$ is the invariant mass. A constant of
motion $P^+$
may be replaced by its eigenvalue  ${\cal P}$ for our states \twostate,
\onestate, and \zerostate. This simple equation leads to complicated
coupled integral eigenvalue equations for wave functions. They are
collected in appendix B. In the following, we discuss the crudest
Tamm-Dancoff truncation (keeping only 2-body states) for the purpose of
illustration. For the isotriplet, the Einstein-Schr\"odinger equation \es\
becomes
\eqn\twotri{
\eqalign{
&{M^2\over 2}\psi_2(x,1-x)
=({m^2\over2}-{e^2\over2\pi})({1\over x}+{1\over1-x})\psi_2(x,1-x)\cr
&-{e^2\over2\pi}\int^1_0dy{\psi_2(y,1-y)\over(x-y)^2}.\cr
}
}
This is the same as that obtained by 'tHooft\ref\thooft{G. 'tHooft, Nucl.
Phys. {\bf
B75} (1974) 461}\ref\oneovern{S. Coleman, in {\sl Pointlike Structure Inside
and
Outside Hadrons} (Plenum Publishing Co.,New York, 1982)} in his study of
$QCD_2$ in
the $1/N$ expansion. It represents a pion consisting of a quark and an
antiquark
interacting through a linear potential.  By integrating over $x$, one gets
\eqn\intgrl{
M^2\int_0^1 dx\psi_2(x,1-x) =m^2\int_0^1 dx({1\over
x}+{1\over1-x})\psi_2(x,1-x).
}
One may easily see that, in the massless limit, it has an eigenvalue $M^2=0$
. On
the other hand, from the equation for the isosinglet,
\eqn\twosing{\eqalign{
&{M^2\over 2}\psi_2'(x,1-x)
=({m^2\over2}-{e^2\over2\pi})({1\over x}+{1\over1-x})\psi_2'(x,1-x)\cr
&+{e^2\over2\pi}\int^1_0dy\psi_2'(y,1-y)(2-{1\over(x-y)^2}),\cr
}
}
one sees that $M^2=\sqrt2e^2/\pi$ is an eigenvalue in the massless limit. Eta
is
heavier than pion. The difference comes from ``2'' in eq.\twosing, which is
due to the annihilation force (the last term of $P_0^-$). In the next
section, we will show that if we include 4-body states there appears a
lighter isosinglet state than eta.

\newsec{Numerical method and results}
\subsec{Basis functions}
There are several ways to discretize the coupled integral eigenvalue
equations\epb\ref\mahiller{ Y. Ma and J. R. Hiller, J. Comp. Phys. {\bf 82}
(1989)
229}. Among them, we think that the method of basis functions\moperry\ is
most
appropriate for our purpose. For the strong coupling, the behavior of the
wave
functions near the edges of momentum region is very important. The DLCQ is
unable to
express this behavior very well, though it is good for moderate values of the
coupling.

The choice of basis
functions is essential for efficient numerical methods. Mo and Perry studied
the
massive Schwinger model by using several basis functions. They showed that
exact
integrability of the matrix elements is very important because otherwise it
would take
much CPU time for numerical integrations for matrix elements. Their
conclusion is
that  Jacobi polynomials are the most appropriate basis functions. When we
includes 4-body states, however, the orthogonality of Jacobi polynomials
does not help us much. In fact their basis functions for 4-body states are
no longer orthogonal. We therefore use simpler basis functions which are
equivalent to their basis functions.

One can easily see by inspection that the 2-body wave functions must vanish
at $x=0$
and $x=1$ because of the mass term. (Here $x$ is a
fraction of momentum. See Appendix B.)  Bergknoff\berg\ showed
that they behave as $x^\beta$, where $\beta$ is the solution of the following
equation, \eqn\mkbeta{
m^2-1+\pi\beta\cot(\pi\beta)=0.
}
(Here and hereafter, we work in the unit $e/\sqrt\pi=1$. The strong coupling
limit
is thus equivalent to the massless limit($m\rightarrow0$).)
 We expand 2-body wave functions in terms of
$f_k(x)\quad (k=0,1,2,\cdots)$,
\eqn\two{
\psi(x,1-x)=\sum_{k=0}^{N_2}a_kf_k(x),
}
where
\eqn\f{
f_k(x)=\left\{\eqalign{ &[x(1-x)]^{\beta+k} \cr
                        &[x(1-x)]^{\beta+k}(2x-1) \cr
                       }.
        \right.
}
The 4-body wave functions (except for $\psi_4$) must also vanish as
$x_i^\beta$ at
$x_i=0\quad (i=1,2,3,4)$. We expand 4-body wave functions in terms of $G_{\bf
k}(x_1,x_2,x_3,x_4)$,
\eqn\four{
\psi(x_1,x_2,x_3,x_4)=\sum_{\bf k}^{N_4}b_{\bf k}G_{\bf
k}(x_1,x_2,x_3,x_4),\quad
\sum_{i=1}^4x_i=1,
}
where
\eqn\G{
G_{\bf k}(x_1,x_2,x_3,x_4)=\left\{\eqalign{
&(x_1x_2x_3x_4)^\beta (x_1-x_2)^{k_1}(x_{12}x_{34})^{k_2}(x_3-x_4)^{k_3}\cr
&(x_1x_2x_3x_4)^\beta
(x_1-x_2)^{k_1}(x_{12}x_{34})^{k_2}(x_3-x_4)^{k_3}x_{12}\cr
}.\right.
}
where $x_{12}=x_1\!+\!x_2$ and $x_{34}=x_3\!+\!x_4$. One can easyly show
that
$(x_1x_2x_3x_4)^\beta x_1^{n_1}x_2^{n_2}x_3^{n_3}x_4^{n_4}$ can be expressed
in terms of $G_{\bf k}(x_1,x_2,x_3,x_4)$ of \G\ uniquely, because of
$x_{12}+x_{34}=1$. According to the symmetries under exchanges,
$k_1$ and
$k_3$ may take only odd or even integers. For example, in the expansion of
$\psi^A$,
$k_1$ and
$k_3$ take only odd values.

\subsec{Eigenvalue equations}
By inserting the expansions of wave functions in terms of basis functions and
projecting the equations to 2- and 4-body basis functions, we obtain
eigenvalue
equations of matrix form. For example, for isosinglet states, we have
\eqn\singl{
\eqalign{
\psi_2'(x,1-x)=&\sum_{k=0}a_kf_k(x),\cr
\psi_3(x_1,x_2,x_3,x_4)=&\sum_{\bf k}b_{\bf k}G_{\bf k}(x_1,x_2,x_3,x_4)
\quad k_1, k_3\ {\rm odd},\cr
\psi_0(x_1,x_2,x_3,x_4)=&\sum_{\bf k}c_{\bf k}G_{\bf k}(x_1,x_2,x_3,x_4)
\quad k_1, k_3\ {\rm even},\cr
}
}
and
\eqn\singmat{
\eqalign{
&M^2\pmatrix{A&0&0\cr0&B&0\cr0&0&B\cr}\pmatrix{a\cr b\cr c\cr}\cr
&=\pmatrix{(m^2-1)C+D&\sqrt6(\tilde E-E)&-\sqrt2(\tilde E-E)\cr
          \sqrt6(\tilde E-E)&{(m^2-1)Q+R+S\atop-4T+6U}&-2\sqrt3 U\cr
           -\sqrt2(\tilde E-E)&-2\sqrt3 U&{(m^2-1)Q+R+S\atop-4T+2U}\cr}
      \pmatrix{a\cr b\cr c\cr},\cr
}
}
where
\eqn\elements{
\eqalign{
A_{kl}=&\int_0^1dx f_k(x)f_l(x),\cr
B_{\bf kl}=&\int_{(4)}G_{\bf k}(x_1,x_2,x_3,x_4)G_{\bf
l}(x_1,x_2,x_3,x_4),\cr
C_{kl}=&\int_0^1dx{f_k(x)f_l(x)\over x(1-x)},\cr
D_{kl}=&\int_0^1dx dy f_k(x)\big(2-{1\over(x-y)^2}\big)f_l(y),\cr
E_{k{\bf l}}=&\int_{(4)}f_k(x_1){1\over(x_2+x_3)^2}G_{\bf
l}(x_1,x_2,x_3,x_4),\cr
\tilde E_{k{\bf l}}=&\int_{(4)}
f_k(1-x_4){1\over(x_2+x_3)^2}G_{\bf l}(x_1,x_2,x_3,x_4),\cr
Q_{\bf kl}=&\int_{(4)}
 G_{\bf k}(x_1,x_2,x_3,x_4)\sum_{i=1}^4{1\over x_i}G_{\bf
l}(x_1,x_2,x_3,x_4),\cr
R_{\bf kl}=&\int_{(6)}
 G_{\bf k}(x_1,x_2,x_3,x_4){1\over(x_1-y_1)^2}G_{\bf l}(y_1,y_2,x_3,x_4),\cr
S_{\bf kl}=&\int_{(6)'}
 G_{\bf k}(x_1,x_2,x_3,x_4){1\over(x_3-y_3)^2}G_{\bf l}(x_1,x_2,y_3,y_4),\cr
T_{\bf kl}=&\int_{(6)''}
 G_{\bf k}(x_1,x_2,x_3,x_4){1\over(x_1-y_1)^2}G_{\bf l}(y_1,x_2,x_3,y_4),\cr
U_{\bf kl}=&\int_{(6)''}
 G_{\bf k}(x_1,x_2,x_3,x_4){1\over(x_1+x_4)^2}G_{\bf l}(y_1,x_2,x_3,y_4),\cr
}
}
with
\eqn\inte{
\eqalign{
\int_{(4)}\equiv&\int_0^1\prod_{i=1}^4dx_i\delta(\sum_{i=1}^4x_i-1),\cr
\int_{(6)}\equiv&\int_0^1\prod_{i=1}^4dx_idy_1dy_2\delta(\sum_{i=1}^4x_i-1)
\delta(x_1+x_2-y_1-y_2),\cr
\int_{(6)'}\equiv&\int_0^1\prod_{i=1}^4dx_idy_3dy_4\delta(\sum_{i=1}^4x_i-1)
\delta(x_3+x_4-y_3-y_4),\cr
\int_{(6)''}\equiv&\int_0^1\prod_{i=1}^4dx_idy_1dy_4\delta(\sum_{i=1}^4x_i-1)
\delta(x_1+x_4-y_1-y_4).\cr
}
}
These are calculated by using the formulae collected in Appendix C. The
matrix
eigenvalue equations for the isotriplet and the isoquintet are presented in
Appendix D.

The matrix eigenvalue equation has the following form containing a ``norm''
$A$,
\eqn\eigeq{
H{\bf x}=EA{\bf x}
}
where $H$ and $A$ are $N\times N$ matrices. By first
solving the eigenvalue problem for the norm $A$,
\eqn\eigA{
A{\bf v}_i=\lambda_i{\bf v}_i,\quad ({\bf v}_i,{\bf v}_j)=\delta_{ij},
}
and then by using the rescaled eigenvectors
${\bf w}_i={\bf v}_i/\sqrt{\lambda_i}$, we can transform \eigeq\ to the
usual form.
\eqn\eigeqq{
W^THW{\bf y}=E{\bf y},
}
where
\eqn\W{
W=({\bf w}_1,\cdots,{\bf w}_N),
}
and ${\bf x}=W{\bf y}$.
We diagonalize this rotated $H$ numerically. Note that the eigenvectors ${\bf
x}_i$
satisfy the relation ${\bf x}_i^TA{\bf x}_j=\delta_{ij}$ if ${\bf y}_i$ are
orthonormalized.

In the following we will be concerned with only the strong coupling region
because the structure of the mass spectrum is relatively simple there.

\subsec{2-body Tamm-Dancoff approximation}
\nfig\onefig{The 2-body Tamm-Dancoff approximation for isotriplets. The
lightest five states are shown against the total number
of basis functions. The convergence is faster for smaller masses. The lowest
state has a very small mass. We call this ``pion''.}
\nfig\zerofig{The 2-body Tamm-Dancoff approximation for isosinglets. The
lightest five states are shown. Note that the spectrum is very similar to
that of the one-flavor model. We call the lowest state ``eta''.}
Let us briefly summerize the results for the crudest
Tamm-Dancoff approximation, keeping only 2-body states.

We have three parameters: the fermion mass (in units of
$e/\sqrt\pi$), $m$, the largest value for $k$ in the expansion \two,
$N_2$,
and the largest value for $k_i$ in the expansion \four, $N_4$,
which is neglected in this subsection\ref\ftwo{The
definitions of $N_2$ and $N_4$ are different from those of Mo and Perry.}.
First of all, we have to know
how many basis functions are necessary in order to produce sufficiently
accurate values.
 Fig.\xfig\onefig\ is the result for the lightest isotriplets. Remarkably,
even only one basis function is enough to get sufficiently good values for
the mass eigenvalues. The lowest mass at
$m=1.0\times10^{-3}$ is $6.02593\times10^{-2}$ for one basis function, and
$6.02593\times10^{-2}$ for 8 basis functions. This should be called
``pion'', whose mass vanishes in the massless limit.
Fig.\xfig\zerofig\ shows the same result for the lightest
isosinglets. The lowest mass at
$m=1.0\times10^{-3}$ is 1.41550 for one basis function, and 1.41550
for 8 basis functions. As we have seen, it should become $\sqrt2$ in the
massless limit. We call this state ``eta'', the lightest isosinglet in
the valence state.
For both cases
the convergence is very fast even for higher states. The differences between
the values for one basis function and for 8 basis functions becomes
appreciable for larger fermion mass, but they are the same in the first five
digits even for
$m=1.0$.

In the next subsection we will include 4-body states keeping $N_2=3$ which
seems large enough.

\subsec{2- and 4-body Tamm-Dancoff approximation}
\nfig\onefigg{The 2- and 4-body Tamm-Dancoff approximation for
isotriplets. The lightest five states are shown with the total number
of 4-body basis functions. Fig. 3a is for $m=1.0\times10^{-3}$ and fig. 3b
for $m=1.0\times10^{-2}$. The increasement of the number of basis functions
leads to finer resolution. Some (``spurious'') states appear when the number
of basis functions is large enough. We think that they are scattering states.
They do not appear to reach the asymptotic values within our calculations.}
\nfig\zerofigg{The 2- and
4-body Tamm-Dancoff approximation for isosinglets. The lightest five states
are shown with the total number of 4-body basis functions. Fig. 4(a) is
for $m=1.0\times10^{-3}$ and fig. 4(b) for $m=1.0\times10^{-2}$. Compare with
\zerofig. The lightest state is not eta, but tends to massless in the
$m\rightarrow0$ limit. There are also ``spurious'' states.}
\nfig\trispec{The isotriplet mass spectrum below $M=15$ with
$(N_2,N_4)=(3,3)$ is
shown in fig. 5(a). Fig. 5(b) is the same result for the isosinglet. Here
$\log$
is the logarithm to the base 10.}
\nfig\triplfig{The low-lying isotriplet states. The pion-pion and
the pion-eta thresholds are shown. We do not find any candidates for
pion-pion, pion-eta bound states but only pion-pion scattering states.}
\nfig\logfig{The relation between the fermion mass and the pion mass. The
straight line is obtained by fitting the first six data by the method of
least
squares.}
\nfig\singlfig{The low-lying isosinglet states. The pion state, the
pion-pion and the eta-eta thresholds are shown for convenience. It is
clear that the lowest state is below the threshold for a wide range of the
coupling constant.}
\nfig\logfigg{The relation between the fermion mass and the eta mass. The
straight line is obtained by fitting the first six data by the method of
least
squares.}
\nfig\fpaifig{The pion decay constant $f_\pi$ as a function of the fermion
mass. It is almost independent of the fermion mass. The straight line is
the least-squares fitting.}
\nfig\twofigg{The
4-body Tamm-Dancoff approximation for isoquintet states. The first five
states are shown with the total number of 4-body basis functions. Fig.
11(a) is for  $m=1.0\times10^{-3}$ and fig. 11(b) for $m=1.0\times10^{-2}$.}
\nfig\quintspec{The isoquintet mass spectrum with $N_4=3$.}
\nfig\quintfig{The low-lying isoquintet states. The pion state and
the pion-pion threshold are shown for convenience. We do not find any
candidates for bound states.}
The convergence of the mass spectrum is not so drastic when the number of
4-body basis functions changes. This is presumably because of the
complicated shapes of the 4-body wave functions. Fig.\xfig\onefigg\ shows
for isotriplets how the masses converge when $N_4$ increases. The total
number of 4-body basis functions is
$N_A+N_{1S}+N_{2S}=2\Big[{N_4+1\over2}\Big]^2\times(N_4+1)
+2\big(2\times\Big[{N_4+1\over2}\Big]\times\Big[{N_4+2\over2}\Big]
\times(N_4+1)\big)$,
where $[\quad]$ is the Gauss' symbol. For $N_4=3$, it is $96$.
Fig.\xfig\zerofigg\ is for isosinglets. The isotriplet and isosinglet mass
spectra
are shown in
\trispec.

{}From these, we see that $N_4=3$ is large enough. In the following, most
calculations are done with $N_4=3$.


The lightest state is pion. Compare with \onefig. The
pion mass at
$m=1.0\times10^{-3}$ is
$5.52473\times10^{-2}$ for $N_4=4$, and it is $5.53050\times10^{-2}$
for $N_4=1$. See \triplfig. The fact that the value does not change
much  (only $9\%$) by the inclusion of 4-body states implies that this
state is in its valence state. In fact, one can see by examining the
wave functions that the probability of being in the 2-body (symmetric
under $x_1\leftrightarrow x_2$)  state is 98.36 percent at
$m=1.0\times10^{-3}$. This state is G-parity  even. If we could
confirm that it is parity odd, our identification would be complete, i.e.,
$1^{-+}$. But in the light-cone quantization, parity is very difficult
to implement. In more than 2 spatial dimensions one may define a kind of
parity operation which leaves the quantization plane intact, by using a
spatial rotation. But in 1 spatial dimension, there are no rotations. In
conclusion, we fail to implement parity in our scheme. It turns out,
however, that charge conjugation and the probabilities of being in
specific states are, in most cases, very powerful in identifing states.

It is interesting to see how the pion mass varies with the fermion
mass.
 Fig.\xfig\logfig\ shows that the pion mass is roughly proportional to the
squre
root of the fermion mass,
\eqn\loglog{
\ln m_\pi=0.564(2)+0.5007(2)\ln m.
}
Note that it seems consistent with the usual notion of current quark
masses.
Grady\grady\ got $0.58\pm0.10$ instead of $0.5007(2)$, and claimed that it
is consistent with $2/3$. We will discuss this in the next section.

It is also easy to find eta in the spectrum. Its mass at
$m=1.0\times10^{-3}$is
1.4154873 for $N_4=4$ and 1.4154960 for $N_4=1$. See \singlfig.
The 2-body result is extremely close to this value. (The difference is
less than $1\times10^{-3}\%$.) The probability of being in the 2-body
(symmetric under $x_1\leftrightarrow x_2$) is more than 99.999 percent
at $m=1.0\times10^{-3}$. The dependence of the mass on the fermion
mass is shown in \logfigg,
\eqn\loglogg{
\ln (m_\eta-\sqrt2)=0.19(1)+0.993(2)\ln m.
}
Note that $m_\eta-\sqrt2$ is roughly proportional to $m$, in contrast to
the pion mass, $m_\pi \sim m^{1/2}$.

This state is G-parity odd.

It is very interesting to see that eta is no longer the lightest
isosinglet once 4-body states are included. Fig.\xfig\zerofigg\ shows that
the lightest isosinglet is not eta, but it tends to be massless in the
$m\rightarrow 0$ limit. Its mass is $9.73543\times10^{-2}$ at
$m=1.0\times10^{-3}$ for $N_4=4$. The convergence is relatively slow. The
isosiglet/isotriplet mass ratio is $1.762$. This should be compared with
$\sqrt3$ in the strong coupling limit, obtained by Coleman. Note, however,
that the pion/eta mass ratio is
$(5.5247\times10^{-2})/1.41549=3.65\times10^{-2}$, the coupling is not
quite strong. Actually $1.762=\sqrt3+{\cal O}(m_\pi/m_\eta)$. Thus we
think that the above value is consistent with Coleman's result. We
will shortly show that this state is G-parity even.

In order to identify states, it is useful to introduce the following
{\it meson} operators,
\eqn\meson{
\eqalign{
A^\dagger_{11}(k_1,k_2)=&b^\dagger_1(k_1)d_2^\dagger(k_2),\cr
A^\dagger_{10}(k_1,k_2)=&-{1\over\sqrt2}[b^\dagger_1(k_1)d_1^\dagger(k_2)
          -b^\dagger_2(k_1)d_2^\dagger(k_2)],\cr
A^\dagger_{1-1}(k_1,k_2)=&-b^\dagger_2(k_1)d_1^\dagger(k_2),\cr
A^\dagger_{00}(k_1,k_2)=&{1\over\sqrt2}[b^\dagger_1(k_1)d_1^\dagger(k_2)
         +b^\dagger_2(k_1)d_2^\dagger(k_2)].\cr
}
}
The isosinglet state $\vert0,0>$ is rewritten in terms of the meson
operators as \eqn\zeromeson{
\eqalign{
\vert 0,0>=&\int^{\cal P}_0{dk_1dk_2\over2\pi\sqrt{k_1k_2}}
            \delta(k_1+k_2-{\cal P})\cr
           &\times\psi_2'(k_1,k_2)A^\dagger_{00}(k_1,k_2)\vert0>\cr
             &+{1\over2\sqrt3}\int^{\cal P}_0{\prod^4_i
dk_i\over(2\pi)^2\sqrt{k_1k_2k_3k_4}}
               \delta(\sum^4_{i=1}k_i-{\cal P})\cr
             &\times\Big\{
              \psi_3(k_1,k_3,k_4,k_2)
              [A^\dagger_{00}(k_1,k_2)A^\dagger_{00}(k_3,k_4)
              -A^\dagger_{00}(k_1,k_4)A^\dagger_{00}(k_3,k_2)]\cr
             &-2\sqrt3\psi_0(k_1,k_3,k_4,k_2)
                 [A^\dagger_{11}(k_1,k_2)A^\dagger_{1-1}(k_3,k_4)\cr
               &-A^\dagger_{10}(k_1,k_2)A^\dagger_{10}(k_3,k_4)
                 +A^\dagger_{1-1}(k_1,k_2)A^\dagger_{11}(k_3,k_4)]
                \Big\}\vert0>.\cr
}
}
{}From this expression it is now clear that $\psi_0$ is the wave
function for the
$\pi$-$\pi$ system. For the lightest isosinglet, the probability of being in
the
4-body $\pi$-$\pi$ state ($\psi_0$) is calculated to be 54.07 percent,
while it is 41.84 percent for the 2-body (anti-symmetric under $x_1
\leftrightarrow x_2$) state. (All the rest is for $\eta$-$\eta$
component.) This state is G-parity even. (See
\charge.)

Now we have obtained the answer to Coleman's question: (1) The lightest
isosinglet $0^{++}$ is a pion-pion bound state. (2) In the weak coupling
 region, its mass will become almost twice of the pion mass, while
the pion mass and the eta mass will become almost degenerate. This is why
the $0^{++}$ state is far away up from these states in the weak coupling
region. (3) Why it's so light? (a) Because it is a bound state of two
pions. Its existence would be found by nonrelativistic reasoning as
Coleman demonstrated for the one-flavor model. Its mass should be smaller
than $2m_\pi$. (b) The annihilation force makes significant effects only
on the 2-body (symmetric under $x_1\leftrightarrow x_2$)\ref\fthree{This
symmetry is very similar to the usual notion of parity. In the following
we give the parity in the parentheses to make the
comparison with Ref.\coleman\  easy.} component of isosinglets. As Coleman
noticed, it shifts symmetric (parity odd) states up. In fact, by examing
the wave functions of isosinglets lighter than eta, one can see that they
do not have the 2-body component which is symmetric under
$x_1\leftrightarrow x_2$ (parity odd). What prevented Coleman from
saying that this is the reason is that the story seems different for
the one-flavor model. The point is that in the one-flavor
model there are no light valence states; the meson has the mass
$e/\sqrt\pi$ in the strong coupling limit, while in the two-flavor model
there is a light valence state (pion). The isosinglet sector knows the
existence of the light isotriplet! Eq.\zeromeson\  shows why the
one-flavor model does not have light states. The information of the light
state is encoded through $\psi_0$, while the 4-body wave function in the
one-flavor model is similar to $\psi_3$. If we
had put $\psi_0=0$ by hand, we would have obtained the spectrum very
similar to that of the one-flavor model.

Let us look for other bound states for the strong couplings in \triplfig\
and \singlfig. We look for eta-eta bound states near the threshold, but such
states cannot be found below the threshold. Above the threshold, we
find a state which may be regarded as a scattering state. (This is
supported by the fact that the $\psi_3$ component is dominant for this
state and that for other states below the threshold the $\psi_0$ and/or
componets
are dominant. For the followings, analogously.)  We look for pi-eta bound
states
in the isotriplet spectrum. We cannot find any candidate, but above the
threshold, we find a state which may be a scattering state. The pi-pi bound
states in the isotriplet are not found, either.

Finally let us discuss isoquintets. Because there are no 2-body wave
functions, we could not determine analytically how the 4-body wave function
behaves near the edge of momenta. We have to vary the $\beta$ to find the
best trial functions. Interestingly, however, for the small fermion mass the
$\beta$ for the best trial functions is just what we obtain by using
\mkbeta, though the difference becomes appreciable for large values. The
spectrum with such $\beta$ is shown with the total number of 4-body
basis functions in
\twofigg. Because the masses for low-lying states are the same for $N_4=3$
and $N_4=4$, the following calculations are done with $N_4=3$.

Fig. \xfig\quintspec\ shows the mass spectrum for various values of the
fermion mass.

The low-lying states are shown in \quintfig. The lightest state is above
the pion-pion threshold, indicating that there are no isoquintet bound
states.

\subsec{Pion decay constant}
Let us first normalize the pion state in the Lorentz-invariant way,
\eqn\normal{
<\pi^a(p)\vert\pi^b(q)>=\delta^{ab}2p^+(2\pi)\delta(p^+-q^+).
}
The pion decay constant may be defined by analogy as
\eqn\fpai{
<0\vert\partial_\mu j^{a\mu}_5(0)\vert \pi^b(p)>=m^2_\pi f_\pi\delta^{ab}.
}
Note that in two dimensions $f_\pi$ is a dimensionless
constant.

We define $\vert\pi^1({\cal P})>=i\sqrt{2\pi}(\vert1,1>-\vert1,-1>)$ which
satisfies
the normalization condition \normal, assuming
\eqn\psisqare{
\eqalign{
&\int_0^1dx_1dx_2\delta(x_1+x_2-1)|\psi_2|^2\cr
&+\int_0^1dx_1dx_2dx_3dx_4\delta(\sum_{i=1}^4x_i-1)
\Big(|\psi^A|^2+|\psi^{1S}|^2+|\psi^{2S}|^2\Big)=1.\cr
}
}
The (phase) factor $i$ has been
introduced for later convenience. Now, from \suc\ and \onestate, one can
easily
obtain
\eqn\divav{
<0\vert\partial_\mu j^{a\mu}_5(0)\vert \pi^b(p)>
=\delta^{ab}{m^2\over\sqrt{2\pi}}\int^1_0dx{\psi_2(x,1-x)\over x(1-x)},
}
thus
\eqn\pdc{
f_\pi={m^2\over\sqrt{2\pi}m_\pi^2}\int^1_0dx{\psi_2(x,1-x)\over x(1-x)}.
}
The right-hand side can be calculated numerically, as shown in
\fpaifig. In the strong
coupling region, it is almost independent of the fermion mass, indicating
that PCAC is a valid concept. In the limit, it is 0.3945,
\eqn\fp{
f_\pi=0.394518(4)+2.26(8)\times10^{-2}m.
}

\newsec{Discussions}
We have obtained the mass spectrum of the massive Schwinger model
numerically  in the LFTD approximation and have seen that the LFTD approach
is very powerful in the study of bound states. In particular, because we
obtain the wave functions simultaneously, the identification of the states is
easy. We also examine PCAC by seeing if the ``pion decay constant'' is
really a constant in the small mass region and how the pion mass changes
with the fermion mass. Remarkably, PCAC is very good even for this
two-dimensional toy model.

In the following we list several unsolved problems.

\item{(1)} The relation between the pion mass and
the fermion mass is consistent with the usual notion of current quark
masses and the pion decay constant depends only weakly on the fermion mass
in the strong coupling region. On the other
hand, however, if we use the current algebra relation,
\eqn\problem{
f_\pi^2m_\pi^2=m<0\vert \sum_{i=1}^2\bar\psi_i\psi_i\vert0>,
}
it follows that the condensation
$<0\vert\sum_{i=1}^2\bar\psi_i\psi_i\vert0>$ is independent of $m$. (Note
that unfortunately we are not able to calculate it directly in our scheme
because we normal-order the fermion bilinear and discard the zero modes
completely.) On the other hand, since it is naively\ref\ffour{ We said
``naively'' because there is a subtlety; it is not clear to us what kind of
``normal-ordering'' their WKB calculations correspond to. The coefficient of
the cosine term of the sine-Gordon Hamiltonian depends on the mass with
respect to which it is normal-ordered. We have found no discussions on this
point in the literature.} expected that
the pion mass is proportional to $m^{2/3}$ in the strong
coupling limit, from the work by Dashen, Hasslacher and
Neveu\ref\dhn{R. F. Dashen, B. Hasslacher and A. Neveu, Phys.
Rev. {\bf D11} (1975) 3424} and Coleman\coleman. It however seems to us
that the value obtained by Grady is also consistent with $m^{1/2}$.
On the other hand, Grady\grady\ also showed that the condensation behaves
like
$m^{0.32\pm0.02}\simeq m^{1/3}$, consistent with the massless case,
where it should be zero because spontaneous symmetry breaking cannot ocuur.
At present, we are not able to point out the reason for this discrepancy.

\item{(2)} It is difficult to implement parity in our formulation. At the
present, we do not know how to do it.

\item{(3)} It is interesting to investigate effects of the non-zero
vacuum angle. One could calculate numerically how $\theta$ affects the
spectrum. The case of $\theta=\pi$ is of particular interest because
Coleman\coleman showed that in this case there appear half-asymptotic
particles. There is a lattice study\ref\hkcm{C. J. Hamer, J. Kogut, D. P.
Crewther and M. M. Mazzolini, Nucl. Phys. {\bf B208} (1982) 413} which
finds the evidence for a phase transition.

\item{(4)} The inclusion of 6-body states will reveal the existence (or
non-existence) of a bound state of 3 pions. One may infer its existence
from the analysis for the one-flavor model when $\theta=0$. It might be
simpler to use DLCQ for this purpose.

\item{(5)} Because we know the (sufficiently accurate) wave functions of
bound states, we may formulate scattering of a bound state off a bound
state. Although the scattering in one spatial dimension appears trivial,
we may study the off-shell physics of the model.


\bigbreak\bigskip
\centerline{{\bf Acknowledgements}}\nobreak
K. H. is grateful to T. Fujita for discussions, especially for the suggestion
that the lightest isosinglet might be a bound state of two pions. This work
is supported by Grant-in-Aid for Encouragement of Young Scientists
(No.05740181) from the Ministry of Education, Science and Culture.


\appendix{A}{Notation and conventions}
In this Appendix, we summerize notation and conventions used in this paper.
They are
essentially the same as those by Perry and Harindranath\ref\hp{
R. J. Perry and A. Harindranath, Phys. Rev. {\bf D43} (1991) 4051}. The
metric is
given by
\eqn\metric{
g^{\mu\nu}=\pmatrix{1&0\cr0&-1\cr}\quad (\mu,\nu=0,1)
\quad g^{\mu\nu}=\pmatrix{0&1\cr1&0\cr} \quad (\mu,\nu=+,-)
}
where
\eqn\x{
x^{\pm}=(x^0\pm x^1)/\sqrt{2}.
}
We treat $x^+$ as our ``time''.

Accordingly, gamma matrices are defined as follows;
\eqn\gammama{\eqalign{
&\quad\gamma^0=\pmatrix{0&1\cr1&0\cr},
\quad\gamma^1=\pmatrix{0&-1\cr1&0\cr},\cr
&\quad\gamma^5=\gamma^0\gamma^1=\pmatrix{1&0\cr0&-1\cr},
\gamma^+=\pmatrix{0&0\cr\sqrt{2}&0\cr},
\quad\gamma^-=\pmatrix{0&\sqrt{2}\cr0&0\cr},\cr
&\quad\gamma^+\gamma^-=\pmatrix{0&0\cr0&2\cr},
\quad\gamma^-\gamma^+=\pmatrix{2&0\cr0&0\cr}.\cr
}
}
thus $\psi=(\psi_R,\psi_L)^T$.
The totally antisymmetric tensor $\epsilon^{\mu\nu}$ is defined by
\eqn\epsi{
\epsilon^{01}=\epsilon^{-+}=+1.
}
\appendix{B}{Einstein-Schr\"odinger equations}
We give a complete set of coupled integral eigenvalue equations
obtained by applying the Hamiltonian\ham\ to the states \twostate,\onestate,
and
\zerostate. We have rescaled the total momentum ${\cal P}$ out by changing
variables
to momentum fractions $x_i=k_i/{\cal P}$. The 2-body wave function
$\psi_2(k_1,k_2)$
is replaced by $\psi_2(x_1,x_2)$ but the 4-body wave function
$\psi^A(k_1,k_2,k_3,k_4)$ by
$\psi^A(x_1,x_2,x_3,x_4)/{\cal P}$, and other wave functions analogously. For
the
4-body wave functions,
$\sum_{i=1}^4x_i=1$. We introduce $\Psi$  and $\Phi$ for notational
convenience.
\eqn\psiphi{
\eqalign{
\Psi&=\psi^{1S}+\psi^{2S}+\sqrt2\psi^A,\cr
\Phi&={1\over2}\psi_0-{\sqrt3\over2}\psi_3.\cr
}
}
\leftline{\sl Isospin=2}
\eqn\Itwo{\eqalign{
&{M^2\over2}\psi_4(x_1,x_2,x_3,x_4)
=({m^2\over2}-{e^2\over2\pi})\sum^4_{i=1}{1\over x_i}
\psi_4(x_1,x_2,x_3,x_4)\cr
&+{e^2\over2\pi}\int_0^1 dy_1dy_2\cr
&\quad\times\Big\{\delta(x_1+x_2-y_1-y_2)
\psi_4(y_1,y_2,x_3,x_4)[{1\over2(x_1-y_1)^2}-{1\over2(x_2-y_1)^2}]\cr
&\quad+\delta(x_3+x_4-y_1-y_2)\psi_4(x_1,x_2,y_1,y_2)
[{1\over2(x_3-y_1)^2}-{1\over2(x_4-y_1)^2}]\cr
&\quad+\delta(x_1+x_4-y_1-y_2)\psi_4(y_1,x_2,y_2,x_3){1\over(x_1-y_1)^2}\cr
&\quad-\delta(x_2+x_4-y_1-y_2)\psi_4(y_1,x_1,y_2,x_3){1\over(x_2-y_1)^2}\cr
&\quad-\delta(x_1+x_3-y_1-y_2)\psi_4(y_1,x_2,y_2,x_4){1\over(x_1-y_1)^2}\cr
&\quad+\delta(x_2+x_3-y_1-y_2)\psi_4(y_1,x_1,y_2,x_4){1\over(x_2-y_1)^2}
\Big\}.\cr
}
}
\leftline{\sl Isospin=1}
\eqn\Ionetwo{\eqalign{
&{M^2\over 2}\psi_2(x,1-x)
=({m^2\over2}-{e^2\over2\pi})({1\over x}+{1\over1-x})\psi_2(x,1-x)\cr
&-{e^2\over2\pi}\int^1_0dy{\psi_2(y,1-y)\over(x-y)^2}\cr
&+{e^2\over\pi}\int^1_0dy_1dy_2dy_3{\delta(y_1+y_2+y_3-x)\over(x-y_1)^2}
[\psi^A(y_1,y_2,y_3,1-x)\cr
&+{1\over\sqrt2}(\psi^{1S}(y_1,y_2,y_3,1-x)+\psi^{2S}(y_1,y_2,y_3,1-x))]\cr
&-{e^2\over\pi}\int^1_0dy_2dy_3dy_4{\delta(y_2+y_3+y_4-(1-x))
\over[(1-x)-y_4]^2}
[\psi^A(x,y_2,y_3,y_4)\cr
&+{1\over\sqrt2}(\psi^{1S}(x,y_2,y_3,y_4)+\psi^{2S}(x,y_2,y_3,y_4))],\cr
}
}
\vfill\eject
\eqn\IoneA{\eqalign{
&{M^2\over 2}\psi^A(x_1,x_2,x_3,x_4)
=({m^2\over2}-{e^2\over2\pi})\sum^4_{i=1}{1\over
x_i}\psi^A(x_1,x_2,x_3,x_4)\cr
&+{e^2\over4\pi}\Big\{\psi_2(x_1,1-x_1)[{1\over(x_2+x_4)^2}
-{1\over(x_2+x_3)^2}]\cr

&\quad-\psi_2(x_2,1-x_2)[{1\over(x_1+x_4)^2}-{1\over(x_1+x_3)^2}]\cr

&\quad-\psi_2(1-x_3,x_3)[{1\over(x_2+x_4)^2}-{1\over(x_1+x_4)^2}]\cr

&\quad+\psi_2(1-x_4,x_4)[{1\over(x_2+x_3)^2}-{1\over(x_1+x_3)^2}]
                \Big\}\cr
&+{e^2\over2\pi}\int^1_0dy_1dy_2\cr
&\times\Big\{\delta(x_1+x_2-y_1-y_2)\psi^A(y_1,y_2,x_3,x_4)
[{1\over2(x_1-y_1)^2}-{1\over2(x_2-y_1)^2}]\cr
&\quad+\delta(x_3+x_4-y_1-y_2)\psi^A(x_1,x_2,y_1,y_2)
[{1\over2(x_3-y_1)^2}-{1\over2(x_4-y_1)^2}]\cr
&\quad-\delta(x_1+x_4-y_1-y_2)\psi^A(x_2,y_1,y_2,x_3){1\over(x_1-y_1)^2}\cr
&\quad+\delta(x_2+x_4-y_1-y_2)\psi^A(x_1,y_1,y_2,x_3){1\over(x_2-y_1)^2}\cr
&\quad+\delta(x_1+x_3-y_1-y_2)\psi^A(x_2,y_1,y_2,x_4){1\over(x_1-y_1)^2}\cr
&\quad-\delta(x_2+x_3-y_1-y_2)\psi^A(x_1,y_1,y_2,x_4)
{1\over(x_2-y_1)^2}\Big\}\cr
&+{e^2\over2\sqrt2\pi}\int^1_0dy_1dy_2\cr
&\times\Big\{\delta(x_1+x_4-y_1-y_2)
\Psi(x_2,y_1,y_2,x_3){1\over(x_1+x_4)^2}\cr
&\quad-\delta(x_2+x_4-y_1-y_2)\Psi(x_1,y_1,y_2,x_3){1\over(x_2+x_4)^2}\cr
&\quad-\delta(x_1+x_3-y_1-y_2)\Psi(x_2,y_1,y_2,x_4){1\over(x_1+x_3)^2}\cr
&\quad+\delta(x_2+x_3-y_1-y_2)\Psi(x_1,y_1,y_2,x_4){1\over(x_2+x_3)^2}
                          \Big\},\cr
}
}

\eqn\IoneS{\eqalign{
&{M^2\over 2}\psi^{1S}(x_1,x_2,x_3,x_4)
=({m^2\over2}-{e^2\over2\pi})\sum^4_{i=1}{1\over
x_i}\psi^{1S}(x_1,x_2,x_3,x_4)\cr
&+{e^2\over4\sqrt2\pi}\Big\{\psi_2(x_1,1-x_1)
[{1\over(x_2+x_4)^2}-{1\over(x_2+x_3)^2}]\cr

&\quad+\psi_2(x_2,1-x_2)[{1\over(x_1+x_4)^2}-{1\over(x_1+x_3)^2}]\cr

&\quad-\psi_2(1-x_3,x_3)[{1\over(x_2+x_4)^2}+{1\over(x_1+x_4)^2}]\cr

&\quad+\psi_2(1-x_4,x_4)[{1\over(x_2+x_3)^2}+{1\over(x_1+x_3)^2}]
                \Big\}\cr
&+{e^2\over2\pi}\int^1_0dy_1dy_2\cr
&\quad\times\Big\{\delta(x_1+x_2-y_1-y_2)
\psi^{1S}(y_1,y_2,x_3,x_4)[{1\over2(x_1-y_1)^2}
+{1\over2(x_2-y_1)^2}]\cr
&\quad+\delta(x_3+x_4-y_1-y_2)\psi^{1S}(x_1,x_2,y_1,y_2)
[{1\over2(x_3-y_1)^2}-{1\over2(x_4-y_1)^2}]\cr
&\quad+\delta(x_1+x_4-y_1-y_2)\psi^{1S}
(x_2,y_1,y_2,x_3){1\over(x_1-y_1)^2}\cr
&\quad+\delta(x_2+x_4-y_1-y_2)\psi^{1S}
(x_1,y_1,y_2,x_3){1\over(x_2-y_1)^2}\cr
&\quad-\delta(x_1+x_3-y_1-y_2)\psi^{1S}
(x_2,y_1,y_2,x_4){1\over(x_1-y_1)^2}\cr
&\quad-\delta(x_2+x_3-y_1-y_2)\psi^{1S}
(x_1,y_1,y_2,x_4){1\over(x_2-y_1)^2}\Big\}\cr
&+{e^2\over4\pi}\int^1_0dy_1dy_2\cr
&\times\Big\{-\delta(x_1+x_4-y_1-y_2)
\Psi(x_2,y_1,y_2,x_3){1\over(x_1+x_4)^2}\cr
&\quad-\delta(x_2+x_4-y_1-y_2)\Psi(x_1,y_1,y_2,x_3){1\over(x_2+x_4)^2}\cr
&\quad+\delta(x_1+x_3-y_1-y_2)\Psi(x_2,y_1,y_2,x_4){1\over(x_1+x_3)^2}\cr
&\quad+\delta(x_2+x_3-y_1-y_2)\Psi(x_1,y_1,y_2,x_4){1\over(x_2+x_3)^2}
                          \Big\},\cr
}
}

\eqn\IoneSS{\eqalign{
&{M^2\over 2}\psi^{2S}(x_1,x_2,x_3,x_4)
=({m^2\over2}-{e^2\over2\pi})\sum^4_{i=1}{1\over
x_i}\psi^{2S}(x_1,x_2,x_3,x_4)\cr
&+{e^2\over4\sqrt2\pi}\Big\{-\psi_2(x_1,1-x_1)
[{1\over(x_2+x_4)^2}+{1\over(x_2+x_3)^2}]\cr

&\quad+\psi_2(x_2,1-x_2)[{1\over(x_1+x_4)^2}+{1\over(x_1+x_3)^2}]\cr

&\quad+\psi_2(1-x_3,x_3)[{1\over(x_2+x_4)^2}-{1\over(x_1+x_4)^2}]\cr

&\quad+\psi_2(1-x_4,x_4)[{1\over(x_2+x_3)^2}-{1\over(x_1+x_3)^2}]
                \Big\}\cr
&+{e^2\over2\pi}\int^1_0dy_1dy_2\cr
&\times\Big\{\delta(x_1+x_2-y_1-y_2)\psi^{2S}
(y_1,y_2,x_3,x_4)[{1\over2(x_1-y_1)^2}-{1\over2(x_2-y_1)^2}]\cr
&\quad+\delta(x_3+x_4-y_1-y_2)\psi^{2S}
(x_1,x_2,y_1,y_2)[{1\over2(x_3-y_1)^2}+{1\over2(x_4-y_1)^2}]\cr
&\quad+\delta(x_1+x_4-y_1-y_2)\psi^{2S}
(x_2,y_1,y_2,x_3){1\over(x_1-y_1)^2}\cr
&\quad-\delta(x_2+x_4-y_1-y_2)\psi^{2S}
(x_1,y_1,y_2,x_3){1\over(x_2-y_1)^2}\cr
&\quad+\delta(x_1+x_3-y_1-y_2)\psi^{2S}
(x_2,y_1,y_2,x_4){1\over(x_1-y_1)^2}\cr
&\quad-\delta(x_2+x_3-y_1-y_2)\psi^{2S}
(x_1,y_1,y_2,x_4){1\over(x_2-y_1)^2}\Big\}\cr
&+{e^2\over4\pi}\int^1_0dy_1dy_2\cr
&\times\Big\{-\delta(x_1+x_4-y_1-y_2)
\Psi(x_2,y_1,y_2,x_3){1\over(x_1+x_4)^2}\cr
&\quad+\delta(x_2+x_4-y_1-y_2)\Psi(x_1,y_1,y_2,x_3){1\over(x_2+x_4)^2}\cr
&\quad-\delta(x_1+x_3-y_1-y_2)\Psi(x_2,y_1,y_2,x_4){1\over(x_1+x_3)^2}\cr
&\quad+\delta(x_2+x_3-y_1-y_2)\Psi(x_1,y_1,y_2,x_4){1\over(x_2+x_3)^2}
                          \Big\}.\cr
}
}
\leftline{\sl Isospin=0}
\eqn\Izerotwo{\eqalign{
&{M^2\over 2}\psi_2'(x,1-x)
=({m^2\over2}-{e^2\over2\pi})({1\over x}+{1\over1-x})\psi_2'(x,1-x)\cr
&+{e^2\over2\pi}\int^1_0dy\psi_2'(y,1-y)(2-{1\over(x-y)^2})\cr
&+{e^2\over2\pi}\int^1_0dy_1dy_2dy_3{\delta(y_1+y_2+y_3-x)\over(x-y_1)^2}\cr
&\times[\sqrt6\psi_3(y_1,y_2,y_3,1-x)-\sqrt2\psi_0(y_1,y_2,y_3,1-x)]\cr
&-{e^2\over2\pi}\int^1_0dy_2dy_3dy_4
{\delta(y_2+y_3+y_4-(1-x))\over[(1-x)-y_4]^2}\cr
&\times[\sqrt6\psi_3(x,y_2,y_3,y_4)-\sqrt2\psi_0(x,y_2,y_3,y_4)],\cr
}
}
\vfill\eject
\eqn\Izerothree{\eqalign{
&{M^2\over 2}\psi_3(x_1,x_2,x_3,x_4)
=({m^2\over2}-{e^2\over2\pi})\sum^4_{i=1}{1\over
x_i}\psi_3(x_1,x_2,x_3,x_4)\cr
&+{\sqrt3e^2\over4\sqrt2\pi}\Big\{

\psi_2'(x_1,1-x_1)[{1\over(x_2+x_4)^2}-{1\over(x_2+x_3)^2}]\cr

&\quad-\psi_2'(x_2,1-x_2)[{1\over(x_1+x_4)^2}-{1\over(x_1+x_3)^2}]\cr

&\quad-\psi_2'(1-x_3,x_3)[{1\over(x_2+x_4)^2}-{1\over(x_1+x_4)^2}]\cr

&\quad+\psi_2'(1-x_4,x_4)[{1\over(x_2+x_3)^2}-{1\over(x_1+x_3)^2}]
                \Big\}\cr
&+{e^2\over2\pi}\int^1_0dy_1dy_2\cr
&\Big\{\delta(x_1+x_2-y_1-y_2)\psi_3(y_1,y_2,x_3,x_4)
[{1\over2(x_1-y_1)^2}-{1\over2(x_2-y_1)^2}]\cr
&\quad+\delta(x_3+x_4-y_1-y_2)\psi_3(x_1,x_2,y_1,y_2)
[{1\over2(x_3-y_1)^2}-{1\over2(x_4-y_1)^2}]\cr
&\quad-\delta(x_1+x_4-y_1-y_2)\psi_3(x_2,y_1,y_2,x_3){1\over(x_1-y_1)^2}\cr
&\quad+\delta(x_2+x_4-y_1-y_2)\psi_3(x_1,y_1,y_2,x_3){1\over(x_2-y_1)^2}\cr
&\quad+\delta(x_1+x_3-y_1-y_2)\psi_3(x_2,y_1,y_2,x_4){1\over(x_1-y_1)^2}\cr
&\quad-\delta(x_2+x_3-y_1-y_2)\psi_3(x_1,y_1,y_2,x_4)
{1\over(x_2-y_1)^2}\Big\}\cr
&+{\sqrt3e^2\over2\pi}\int^1_0dy_1dy_2
\Big\{-\delta(x_1+x_4-y_1-y_2)\Phi(x_2,y_1,y_2,x_3){1\over(x_1+x_4)^2}\cr
&\quad+\delta(x_2+x_4-y_1-y_2)\Phi(x_1,y_1,y_2,x_3){1\over(x_2+x_4)^2}\cr
&\quad+\delta(x_1+x_3-y_1-y_2)\Phi(x_2,y_1,y_2,x_4){1\over(x_1+x_3)^2}\cr
&\quad-\delta(x_2+x_3-y_1-y_2)\Phi(x_1,y_1,y_2,x_4){1\over(x_2+x_3)^2}
                          \Big\},\cr
}
}
\eqn\Izerozero{
\eqalign{
&{M^2\over 2}\psi_0(x_1,x_2,x_3,x_4)
=({m^2\over2}-{e^2\over2\pi})\sum^4_{i=1}{1\over
x_i}\psi_0(x_1,x_2,x_3,x_4)\cr
&+{e^2\over4\sqrt2\pi}\Big\{

\psi_2'(x_1,1-x_1)[{1\over(x_2+x_4)^2}+{1\over(x_2+x_3)^2}]\cr

&\quad+\psi_2'(x_2,1-x_2)[{1\over(x_1+x_4)^2}+{1\over(x_1+x_3)^2}]\cr

&\quad-\psi_2'(1-x_3,x_3)[{1\over(x_2+x_4)^2}+{1\over(x_1+x_4)^2}]\cr

&\quad-\psi_2'(1-x_4,x_4)[{1\over(x_2+x_3)^2}+{1\over(x_1+x_3)^2}]
                \Big\}\cr
&+{e^2\over2\pi}\int^1_0dy_1dy_2\cr
&\Big\{\delta(x_1+x_2-y_1-y_2)\psi_0(y_1,y_2,x_3,x_4)
[{1\over2(x_1-y_1)^2}+{1\over2(x_2-y_1)^2}]\cr
&\quad+\delta(x_3+x_4-y_1-y_2)\psi_0(x_1,x_2,y_1,y_2)
[{1\over2(x_3-y_1)^2}+{1\over2(x_4-y_1)^2}]\cr
&\quad-\delta(x_1+x_4-y_1-y_2)\psi_0(x_2,y_1,y_2,x_3){1\over(x_1-y_1)^2}\cr
&\quad-\delta(x_2+x_4-y_1-y_2)\psi_0(x_1,y_1,y_2,x_3){1\over(x_2-y_1)^2}\cr
&\quad-\delta(x_1+x_3-y_1-y_2)\psi_0(x_2,y_1,y_2,x_4){1\over(x_1-y_1)^2}\cr
&\quad-\delta(x_2+x_3-y_1-y_2)\psi_0(x_1,y_1,y_2,x_4)
{1\over(x_2-y_1)^2}\Big\}\cr
&+{e^2\over2\pi}\int^1_0dy_1dy_2
\Big\{\delta(x_1+x_4-y_1-y_2)\Phi(x_2,y_1,y_2,x_3){1\over(x_1+x_4)^2}\cr
&\quad+\delta(x_2+x_4-y_1-y_2)\Phi(x_1,y_1,y_2,x_3){1\over(x_2+x_4)^2}\cr
&\quad+\delta(x_1+x_3-y_1-y_2)\Phi(x_2,y_1,y_2,x_4){1\over(x_1+x_3)^2}\cr
&\quad+\delta(x_2+x_3-y_1-y_2)\Phi(x_1,y_1,y_2,x_4){1\over(x_2+x_3)^2}
                          \Big\}.\cr
}
}

\appendix{C}{ Some formulae}
This is a list of useful integral formulae for calculating matrix elements by
using
the basis functions \f and \G. Principal-value integrals are understood.
\eqn\one{\eqalign{
I^S(\alpha,\beta)=&\int_0^1dxdy{[x(1-x)]^\alpha[y(1-y)]^\beta\over(x-y)^2}\cr
=&-{\alpha\beta\over2(\alpha+\beta)}B(\alpha,\alpha)B(\beta,\beta)\cr
I^A(\alpha,\beta)=&\int_0^1dxdy{[x(1-x)]^\alpha[y(1-y)]^\beta
\over(x-y)^2}(2x-1)(2y-1)\cr
=&-{\alpha\beta\over2(\alpha+\beta)(\alpha+\beta+1)}
B(\alpha,\alpha)B(\beta,\beta)\cr
}
}
\eqn\two{\eqalign{
&\int\prod^4_{i=1}dx_i\delta(\sum_{i=1}^4x_i-1)
(x_2+x_4)^{-2}[x_1(1-x_1)]^{\beta+k}
(x_1x_2x_3x_4)^\beta x_1^{n_1}x_2^{n_2}x_3^{n_3}x_4^{n_4}\cr
&=B(4\beta+k+n_2+n_3+n_4+1,2\beta+k+n_1+1)\cr
&\times B(\beta+n_4+1,\beta+n_2+1)
B(2\beta+n_2+n_4,\beta+n_3+1)\cr
}
}
\eqn\three{\eqalign{
&\int\prod^4_{i=1}dx_i\delta(\sum_{i=1}^4x_i-1)
(x_1x_2x_3x_4)^\beta x_1^{n_1}x_2^{n_2}x_3^{n_3}x_4^{n_4}\cr
&={\Gamma(2\beta+n_1+1)\Gamma(2\beta+n_2+1)\Gamma(2\beta+n_3+1)
\Gamma(2\beta+n_4+1)\over\Gamma(8\beta+n_1+n_2+n_3+n_4+4)}\cr
}
}
\eqn\four{\eqalign{
&\int\prod^4_{i=1}dx_idy_1dy_2\delta(\sum_{i=1}^4x_i-1)
\delta(x_1+x_2-y_1-y_2)\cr
&(x_1+x_2)^{-2}
(x_1x_2x_3x_4)^\beta(y_1y_2)^\beta(x_3x_4)^\beta
x_1^{n_1}x_2^{n_2}x_3^{n_3}x_4^{n_4}y_1^{m_1}y_2^{m_2}\cr
&=B(\beta+m_1+1,\beta+m_2+1)B(\beta+n_1+1,\beta+n_2+1)
\cr
&\times{\Gamma(4\beta+n_1+n_2+m_1+m_2+1)
\Gamma(2\beta+n_3+1)\Gamma(2\beta+n_4+1)\over
\Gamma(8\beta+n_1+n_2+n_3+n_4+m_1+m_2+3)}\cr
}
}
\eqn\five{\eqalign{
&\int\prod^4_{i=1}dx_idy_1dy_2\delta(\sum_{i=1}^4x_i-1)
\delta(x_1+x_2-y_1-y_2)\cr
&(x_1-y_1)^{-2}
(x_1x_2x_3x_4)^\beta(y_1y_2)^\beta(x_3x_4)^\beta
x_1^{n_1}x_2^{n_2}x_3^{n_3}x_4^{n_4}y_1^{m_1}y_2^{m_2}\cr
&={\Gamma(4\beta+n_1+n_2+m_1+m_2+1)
\Gamma(2\beta+n_3+1)\Gamma(2\beta+n_4+1)\over
\Gamma(8\beta+n_1+n_2+n_3+n_4+m_1+m_2+3)}\cr
&\times D(n_1,n_2,m_1,m_2)\cr
}
}
where
\eqn\six{
\eqalign{
&D(n_1,n_2,m_1,m_2)=\int_0^1 dx
dy{x^{\beta+n_1}(1-x)^{\beta+n_2}
y^{\beta+m_1}(1-y)^{\beta+m_2}\over(x-y)^2}\cr
&=2^{-|n_1-n_2|-|m_1-m_2|}\cr
&\times\sum^{|n_1-n_2|}_{k=0\atop even}\sum_{r=0}^{k/2}{|n_1-n_2|\choose
k} {k/2 \choose r}\sum^{|m_1-m_2|}_{l=0\atop
even}\sum_{s=0}^{l/2}{|m_1-m_2|\choose l}
{l/2 \choose s}(-4)^{r+s}\cr
&\times I^S(\beta+\underline{n}+r,\beta+\underline{m}+s)\cr
&+2^{-|n_1-n_2|-|m_1-m_2|}\epsilon(n_1-n_2)\epsilon(m_1-m_2)\cr
&\times\sum^{|n_1-n_2|}_{k=0\atop odd}\sum_{r=0}^{k/2}{|n_1-n_2|\choose
k} {k/2 \choose r}\sum^{|m_1-m_2|}_{l=0\atop
odd}\sum_{s=0}^{l/2}{|m_1-m_2|\choose l}
{l/2 \choose s}(-4)^{r+s}\cr
&\times I^A(\beta+\underline{n}+r,\beta+\underline{m}+s)\cr
}
}
with $\underline{n}=\hbox{\rm min}(n_1,n_2)$ and $\underline{m}=\hbox{\rm
min}(m_1,m_2)$.

\appendix{D}{Matrix eigenvalue equations}
The following are matrix eigenvalue equations for the isotriplet and for the
isoquintet.

\leftline{\sl Isotriplet:}
\eqn\tripl{
\eqalign{
\psi_2(x,1-x)=&\sum_{k=0}a_kf_k(x),\cr
\psi^A(x_1,x_2,x_3,x_4)=&\sum_{\bf k}b_{\bf k}G_{\bf k}(x_1,x_2,x_3,x_4)
\quad k_1, k_3\ {\rm odd},\cr
\psi^{1S}(x_1,x_2,x_3,x_4)=&\sum_{\bf k}c_{\bf k}G_{\bf k}(x_1,x_2,x_3,x_4)
\quad k_1\ {\rm even}\quad , k_3\ {\rm odd},\cr
\psi^{2S}(x_1,x_2,x_3,x_4)=&\sum_{\bf k}d_{\bf k}G_{\bf k}(x_1,x_2,x_3,x_4)
\quad k_1\ {\rm odd}\quad , k_3\ {\rm even}.\cr
}
}

\eqn\trimat{
\eqalign{
&M^2\!\pmatrix{A&0&0&0\cr0&B&0&0\cr0&0&B&0\cr0&0&0&B\cr}
\pmatrix{a\cr b\cr c\cr d\cr}\cr
&=\!\pmatrix{(m^2-1)C+\bar D&2(\tilde E-E)&\sqrt2(\tilde E-E)&\sqrt2(\tilde
E-E)\cr
          2(\tilde E-E)&{(m^2-1)Q+R+S\atop-4T+4U}&-2\sqrt2 U&-2\sqrt2 U\cr
          \sqrt2(\tilde E-E)&-2\sqrt2 U&{(m^2-1)Q+R+S\atop-4T+2U}&2U\cr
           \sqrt2(\tilde E-E)&-2\sqrt2 U&2U&{(m^2-1)Q+R+S\atop-4T+2U}\cr}
      \!\pmatrix{a\cr b\cr c\cr d\cr},\cr
}
}
where
\eqn\dbar{
\bar D_{kl}=-\int_0^1 dxdy{f_k(x)f_l(y)\over(x-y)^2}.
}

\leftline{\sl Isoquintet}
\eqn\quint{
\psi_4(x_1,x_2,x_3,x_4)=\sum_{\bf k}b_{\bf k}G_{\bf k}(x_1,x_2,x_3,x_4)
\quad k_1, k_3\ {\rm odd}.
}
\eqn\quinmat{
M^2Bb=\big((m^2-1)Q+R+S-4T\big)b.
}


\listrefs
\listfigs

\bye